\documentclass[11pt,letterpaper]{article}

\usepackage[utf8]{inputenc}
\usepackage[T1]{fontenc}
\usepackage{microtype}

\usepackage[letterpaper,top=1in,bottom=2cm,left=1in,right=1in]{geometry}

\usepackage[skip=5pt plus 3pt minus 1pt,parfill=0pt]{parskip}
\makeatletter
\def\thm@space@setup{%
  \thm@preskip=\parskip \thm@postskip=0pt
}
\renewcommand\paragraph{\@startsection{paragraph}{4}{\z@}{\parskip}{-1em}{\normalfont\normalsize\bfseries}}
\makeatother

\usepackage{amsmath}
\usepackage{amsthm}
\usepackage{mathtools}

\usepackage{amssymb}
\usepackage{bm,dsfont}
\usepackage{csquotes}
\usepackage{nicefrac}

\usepackage{xcolor}

\definecolor{col1}{HTML}{56b3e9}
\definecolor{col2}{HTML}{e69f00}
\definecolor{col3}{HTML}{009e74}
\definecolor{col4}{HTML}{cc79a7}
\definecolor{col5}{HTML}{d55e00}
\definecolor{col6}{HTML}{0071b2}

\usepackage[small]{caption}
\usepackage{subcaption}
\usepackage{tikz}
\usepackage{pgfplots}
\pgfplotsset{compat=1.18}

\usepackage{booktabs}
\usepackage{array}

\usepackage[shortlabels]{enumitem}

\usepackage[toc,page]{appendix}

\usepackage[
 backend=biber,
 style=alphabetic,
 citetracker,
 hyperref=auto,
 maxcitenames=10,
 minalphanames=3,
 maxalphanames=4,
 sortcites,
 sorting=ynt,
 maxbibnames=12,
 date=year,
 isbn=false,
 url=false,
 doi=false,
 eprint=false,
]{biblatex}
\newbibmacro{string+doiurl}[1]{%
  \iffieldundef{doi}{%
    \iffieldundef{url}{%
      #1
    }{%
      \href{\thefield{url}}{#1}%
    }%
  }{%
    \href{http://dx.doi.org/\thefield{doi}}{#1}%
  }%
}

\AtEveryCitekey{%
  \ifciteseen
    {\defcounter{maxnames}{2}}
    {}%
}

\DeclareFieldFormat%
[article,inbook,incollection,inproceedings,patent,thesis,unpublished]
  {title}{\usebibmacro{string+doiurl}{#1}}

\usepackage[colorlinks]{hyperref}
\hypersetup{citecolor=green!60!black,linkcolor=blue!70!black,urlcolor=blue!70!black}
\usepackage[capitalize, noabbrev, nameinlink]{cleveref}
\crefname{condition}{condition}{conditions}
\Crefname{condition}{Condition}{Conditions}
\crefname{observation}{Observation}{Observations}
\usepackage{xurl}
\hypersetup{breaklinks=true}

\usepackage{xspace}

\allowdisplaybreaks

\newtheorem{theorem}{Theorem}[section]
\newtheorem{lemma}[theorem]{Lemma}

\newtheorem{observation}[theorem]{Observation}

\theoremstyle{definition}
\newtheorem{definition}[theorem]{Definition}

\theoremstyle{remark}

\AddToHook{env/theorem/begin}{\crefalias{theorem}{theorem}}
\AddToHook{env/lemma/begin}{\crefalias{theorem}{lemma}}
\AddToHook{env/claim/begin}{\crefalias{theorem}{claim}}
\AddToHook{env/fact/begin}{\crefalias{theorem}{fact}}
\AddToHook{env/corollary/begin}{\crefalias{theorem}{corollary}}
\AddToHook{env/proposition/begin}{\crefalias{theorem}{proposition}}
\AddToHook{env/observation/begin}{\crefalias{theorem}{observation}}
\AddToHook{env/definition/begin}{\crefalias{theorem}{definition}}
\AddToHook{env/example/begin}{\crefalias{theorem}{example}}
\AddToHook{env/remark/begin}{\crefalias{theorem}{remark}}

\DeclarePairedDelimiter{\abs}{\lvert}{\rvert}

\newcommand{\bone}{\mathbf{1}}
\newcommand{\dd}{\,\mathrm{d}}

\newcommand{\OPT}{{\mathrm{OPT}}}

\newcommand{\eps}{\varepsilon}

\usetikzlibrary{decorations.pathreplacing,patterns}
\newcommand{\E}{\mathbb{E}}
\newcommand{\cD}{\mathcal{D}}
\newcommand{\cA}{\mathcal{A}}
\newcommand{\Amp}{\mathrm{Amp}}
\newcommand{\Avg}{\mathrm{Avg}}

\title{Weighted Flow Time Does Not Admit\\ Constant-Competitive Randomized Algorithms}

\author{Alexander Lindermayr\thanks{Institut für Mathematik, Technische Universität Berlin, Germany (\texttt{alexander.lindermayr@tu-berlin.de}). Supported by the Deutsche Forschungsgemeinschaft (DFG, German Research Foundation) under Germany’s Excellence Strategy -- The Berlin Mathematics Research Center MATH$^+$ (EXC-2046/1, EXC-2046/2, project ID: 390685689).}}
\date{}

\begin{document}

\maketitle

\begin{abstract}
We study online preemptive scheduling on a single machine with the objective of minimizing the total weighted flow time. Bansal and Chan (SODA~2009) showed that no deterministic online algorithm is $O(1)$-competitive for this problem. We show that no randomized online algorithm is $O(1)$-competitive against an oblivious adversary either. This settles the remaining case of the long-standing question whether the problem admits an $O(1)$-competitive online algorithm. Quantitatively, every randomized online algorithm has a competitive ratio in~$\Omega(\min\{\log\log W,\log\log\log\min\{P,D,n\}\})$, where $n$ is the number of jobs and $W$, $P$, and~$D$ are the ratios of the largest to the smallest job weight, size, and density (weight divided by size), respectively. The lower bound also holds with high probability, with the same asymptotic dependence on the instance parameters. For every~$\rho\ge1$ and every randomized online algorithm, there is an instance on which the cost of the algorithm is at least~$\rho$ times the optimal cost with probability at least $1-2^{-2^\rho}$. Both lower bounds hold even if the algorithm knows in advance the size and weight of every job in release order and, for any fixed~$\eps>0$, every interarrival time up to a factor of $1+\eps$.
\end{abstract}

\section{Introduction}

In online scheduling, jobs arrive over time and must be scheduled without knowledge of the future. A fundamental objective is the total weighted flow time, that is, the total time that the jobs spend in the system, weighted by their importance~\cite{PruhsST04}. We study this objective on a single machine with preemption. Each job~$j$ has a release time~$r_j$, a size~$p_j$, and a weight~$w_j$, all of which become known at time~$r_j$. The machine processes at most one job at a time and may interrupt a job and resume it later. If job~$j$ completes at time~$C_j$, its flow time is $C_j-r_j$, and the cost of a schedule is its total weighted flow time $\sum_j w_j(C_j-r_j)$. An online algorithm decides at each time which job to process, knowing only the jobs released so far, and it is \emph{$\rho$-competitive} if its (expected) cost is at most $\rho$ times the optimal cost on every instance.

With unit weights, the online rule that always processes a job with the shortest remaining processing time (SRPT) is optimal~\cite{Schrage68}. Non-uniform weights make the problem much harder. Even the offline problem is strongly NP-hard~\cite{LabetoulleLLR84}, and a polynomial-time approximation scheme~\cite{ArmbrusterRW26} was found only after a long line of work~\cite{ChekuriK02,BansalP14,BatraGK23,FeigeKL19,RohwedderW21}.

For the online problem, the best known upper bounds on the competitive ratio grow logarithmically with the parameters of the instance. Let $W$, $P$, and~$D$ be the ratios of the largest to the smallest job weight, size, and density $w_j/p_j$, respectively. The first nontrivial algorithm, by Chekuri, Khanna, and Zhu~\cite{ChekuriKZ01}, is $O(\log^2P)$-competitive and needs to know~$P$ in advance. Bansal and Dhamdhere~\cite{BansalD07} gave an $O(\log W)$-competitive algorithm. Azar and Touitou~\cite{AzarT18} gave $O(\log P)$- and $O(\log D)$-competitive algorithms and combined them with the algorithm of Bansal and Dhamdhere into an $O(\log\min\{P,D,W\})$-competitive deterministic algorithm.

Pruhs, Sgall, and Torng~\cite{PruhsST04} called the existence of an $O(1)$-competitive algorithm perhaps the most intriguing open question about flow time in online scheduling. As Bansal and Chan~\cite{BansalC09} note, it was widely believed that such an algorithm exists. They showed, however, that every deterministic online algorithm has a competitive ratio in
\begin{equation*}
  \Omega\biggl(\min\biggl\{\sqrt{\frac{\log W}{\log\log W}},\,\sqrt{\frac{\log\log P}{\log\log\log P}}\biggr\}\biggr)\,.
\end{equation*}
Their proof starts from the trivial lower bound of one and improves it step by step, using an adaptive adversary that chooses the next jobs in response to the algorithm's scheduling decisions. By contrast, a randomized algorithm is evaluated against an \emph{oblivious} adversary, which fixes the instance in advance, without knowing the algorithm's random choices. The adaptive construction of Bansal and Chan~\cite{BansalC09} therefore does not apply to randomized algorithms.

A lower bound for deterministic algorithms does not settle the randomized case, because randomization is provably powerful in preemptive online scheduling. For example, when job sizes remain unknown until completion (so-called \emph{nonclairvoyant} scheduling), every deterministic algorithm for unweighted flow time on a single machine has a competitive ratio in $\Omega(n^{1/3})$, where $n$ is the number of jobs~\cite{MotwaniPT94}. Yet a randomized version of the multilevel feedback algorithm~\cite{KalyanasundaramP03} is $O(\log n)$-competitive~\cite{BecchettiL04}, which is best possible up to a constant factor~\cite{MotwaniPT94}.

For weighted flow time on a single machine, the best lower bound for randomized algorithms that we are aware of is $4/3$~\cite{ChekuriKZ01,EpsteinS03}. Thus it has remained open whether a randomized online algorithm can be $O(1)$-competitive.

\subsection{Our Results}
We answer this question negatively.

\begin{theorem}\label{thm:main}
No randomized online algorithm for minimizing weighted flow time on a single machine is $O(1)$-competitive against an oblivious adversary. More precisely, for every $\rho\ge1$, there is a probability distribution~$\cD$ over finitely many instances such that every deterministic online algorithm~$A$ satisfies $\E_{I\sim\cD}\bigl[F_A(I)\bigr]\ge \rho\cdot\E_{I\sim\cD}\bigl[\OPT(I)\bigr]$.
\end{theorem}

Here $F_A(I)$ denotes the weighted flow time of the schedule that~$A$ produces on the instance~$I$, and $\OPT(I)$ is the optimal weighted flow time of~$I$. By Yao's principle~\cite{Yao77}, the second statement implies the first.
Thus, even with randomization, the competitive ratio must grow with the parameters of the instance. Our construction makes this dependence explicit in terms of~$W$, $P$, $D$, and the number~$n$ of jobs. All logarithms are to base~$2$.

\begin{theorem}\label{thm:quant}
For all $W,P,D,n\ge16$, every randomized online algorithm for weighted flow time has a competitive ratio in
\begin{equation*}
  \Omega\bigl(\min\{\log\log W,\,\log\log\log\min\{P,D,n\}\}\bigr)
\end{equation*}
against an oblivious adversary on instances with at most~$n$ jobs and weight, size, and density ratios at most~$W$, $P$, and~$D$, respectively.
\end{theorem}

The minimum reflects that our hard instances simultaneously have large weight, size, and density ratios and many jobs. For the three ratios this cannot be avoided, since the $O(\log\min\{P,D,W\})$-competitive algorithm of Azar and Touitou~\cite{AzarT18} is $O(1)$-competitive as soon as one of them is bounded. If only one parameter is restricted, then~\cref{thm:quant} gives the lower bound $\Omega(\log\log W)$ in terms of the weight ratio and the lower bounds $\Omega(\log\log\log P)$, $\Omega(\log\log\log D)$, and $\Omega(\log\log\log n)$ in terms of the size ratio, the density ratio, and the number of jobs.

Our lower bound also holds with high probability, with the same dependence on the parameters.

\begin{theorem}\label{thm:probability}
There is an absolute constant $C\ge1$ such that, for every $\rho\ge1$ and every randomized online algorithm~$\cA$, there is an instance~$I$ on which~$\cA$ incurs cost at least $\rho\cdot\OPT(I)$ with probability at least $1-2^{-2^\rho}$. This instance has weight ratio at most $2^{2^{C\rho}}$, and its size ratio, density ratio, and number of jobs are at most $2^{2^{2^{C\rho}}}$.
\end{theorem}

None of these lower bounds relies on uncertainty about the jobs themselves. In our hard distributions, the sizes and weights of all jobs are fixed, and only the times between consecutive releases are random, each up to a factor of $1+\eps$ for an arbitrary $\eps>0$.

\begin{theorem}\label{thm:interarrival}
For every $\eps>0$, there is a constant $C\ge1$ such that the following holds for every $\rho\ge1$. There are $n$ jobs with sizes $p_1,\dots,p_n$, weights $w_1,\dots,w_n$, and independent random variables $\gamma_1,\dots,\gamma_{n-1}$ with finitely many values, where each~$\gamma_j$ takes values in $[g_j,(1+\eps)g_j]$ for some $g_j>0$, with the following property. Let~$I$ be the random instance in which the $j$th job has release time $r_j=\gamma_1+\dots+\gamma_{j-1}$. Then every deterministic online algorithm~$A$ satisfies
\begin{equation*}
  \E\bigl[F_A(I)\bigr]\ge\rho\cdot\E\bigl[\OPT(I)\bigr]
  \qquad\text{and}\qquad
  \Pr\bigl[F_A(I)\ge\rho\cdot \OPT(I)\bigr]\ge1-2^{-2^\rho}\,.
\end{equation*}
The weight ratio of~$I$ is at most $2^{2^{C\rho}}$, and its size ratio, density ratio, and number of jobs are at most $2^{2^{2^{C\rho}}}$.
\end{theorem}

The algorithm may depend on~$n$, on the sizes and weights, and on the distributions of the interarrival times $\gamma_j=r_{j+1}-r_j$. Both bounds therefore hold even if all of these are known in advance. The only uncertainty is then a factor of $1+\eps$ in each interarrival time, and hence also in each release time, because every release time is a sum of interarrival times. %

\subsection{Technical Overview}\label{sec:overview}

It suffices to construct, for every $\rho\ge1$, a distribution over instances on which every deterministic online algorithm has expected cost at least $\rho$ times the expected optimal cost. Such a distribution cannot react to the algorithm, so it cannot choose later jobs based on which earlier jobs remain unfinished. Instead, our distributions hide the time at which the algorithm's \emph{active weight}, the total weight of released but uncompleted jobs, is compared with the minimum possible active weight.

\paragraph{Active weight at a hidden time.}
Weighted flow time is the integral of the active weight over time. An algorithm whose active weight is always at most $c$ times the minimum possible active weight at that time is therefore $c$-competitive. Conversely, Becchetti, Leonardi, Marchetti-Spaccamela, and Pruhs~\cite{BecchettiLMP06} showed that every $c$-competitive deterministic online algorithm essentially has this property, as long as nearly finished jobs are not counted, using a stream of short, dense jobs. We use this converse direction. We first create a large expected gap in active weight at a hidden time and then use such a stream to make this gap persist long enough to dominate the weighted flow time. Bansal and Chan~\cite{BansalC09} use the same conversion.

\paragraph{Hard terminal distributions.}
Our building blocks are \emph{terminal distributions} over pairs $(I,T)$ of instances and times, where~$T$ is later than all release times in~$I$. The algorithm sees the jobs of~$I$ as they arrive, but is not given~$T$. Let $w^*(I,T)$ be the minimum active weight at~$T$ among all feasible schedules. Online, we count only \emph{robust} jobs, whose remaining size at~$T$ is at least~$1$. These jobs form the \emph{robust backlog} and cannot finish before~$T+1$. We cap their total weight at~$M$. A terminal distribution is \emph{$(c,M)$-hard} if $\E[w^*(I,T)]\le1$ and every deterministic online algorithm has expected capped robust backlog at least~$c$. We call~$c$ the \emph{gap} and~$M$ the \emph{cap}.

\paragraph{Flooding.}
A $(2\rho,M)$-hard terminal distribution yields the lower bound~$\rho$ by \emph{flooding}. Starting at~$T$, a long stream of jobs of weight~$1/2$ and density~$M$ keeps the machine busy. Until the algorithm has spent one unit of processing on old jobs, its robust backlog remains unfinished. Afterwards, the delayed stream jobs have total weight at least~$M$. Hence the active weight of the algorithm never drops below its capped robust backlog at~$T$ while the stream lasts. An offline schedule can instead leave weight $w^*(I,T)$ unfinished and process each stream job immediately, which keeps its active weight at $w^*(I,T)+1/2$. A sufficiently long stream makes this gap dominate the costs before and after it.

It thus suffices to construct $(c,M)$-hard terminal distributions with arbitrarily large gap~$c$. We start from the trivial $(1,M)$-hard distribution that consists of a single job of size~$2$ and weight~$1$, released at time~$0$, with terminal time~$1$, and we alternate two operations, \emph{amplification} and \emph{averaging}. Both build on the following way of combining copies of a hard distribution.

\paragraph{Copying and scaling.}
We concatenate independent samples of a hard distribution, starting copy~$i$ at the terminal time of copy~$i-1$ and stretching its times and sizes by a factor~$s_i$. Measured at this scale, its robust backlog consists of jobs with remaining size at least~$s_i$. All later copies together have duration and total size less than~$s_i/2$, so this backlog survives them even if an algorithm only works on the backlog and not on the later copies. Conditioned on the earlier copies, copy~$i$ remains a fresh sample, and later copies arrive at or after its terminal time. Thus its capped robust backlog~$X_i$ has conditional expectation at least~$c$, even though the algorithm adapts to the past. An offline schedule can process each copy in its own window, leaving expected weight at most~$1$ per copy.

\paragraph{Amplification.}
We take $k=2M+1$ copies of a $(c,M)$-hard distribution and, immediately after them, release a heavy job~$J$ of weight~$2M$ whose size~$V$ is at least the total size of all copies. The new terminal time is one of $k$ checkpoints $h_k<h_{k-1}<\dots<h_1$ after the release of~$J$, where checkpoint~$h_i$ leaves enough time to complete~$J$ and copies $i+1,\dots,k$. We choose the checkpoint~$h_K$ at random, with $\Pr[K\ge i]=2^{1-i}$ for $i=1,\dots,k$. An offline schedule that completes~$J$ and copies $i+1,\dots,k$ by~$h_i$ leaves only jobs of the other~$i$ copies. Each of these copies leaves expected weight at most~$1$, so the schedule leaves expected weight at most~$i$. At the earliest checkpoint~$h_k$, this bound is~$k$, and it is better to complete all copies and leave only~$J$, of weight $2M=k-1$. The offline schedule therefore leaves expected weight at most~$K$, and one unit less if $K=k$. Averaged over~$K$, the schedule leaves expected weight at most $\E[K]-\Pr[K=k]$ at the terminal time~$h_K$ of the new instance.

The online algorithm faces a dilemma at every checkpoint~$h_i$. If it has nearly finished~$J$ by~$h_i$, it had too little time for the robust backlog of copy~$i$, which then survives at~$h_i$ and at all earlier checkpoints $h_k,\dots,h_{i+1}$. Since~$J$ is then also nearly finished at the later checkpoints $h_{i-1},\dots,h_1$, the algorithm holds the robust backlogs of all copies $1,\dots,i$ at~$h_i$, each of expected capped weight at least~$c$, whereas the offline schedule leaves expected weight at most~$1$ per copy. Otherwise~$J$ itself is still robust at~$h_i$ and counts with its weight~$2M$. In this case, the algorithm may have used the time that it did not spend on~$J$ to clear the robust backlogs of some of the copies $1,\dots,i$. The weight of~$J$ makes up for the cleared backlogs in expectation, because each of them counts for at most its cap~$M$ and the checkpoint probabilities halve with each index. Hence the expected capped robust backlog at the terminal time~$h_K$ of the new instance is at least $c\cdot\E[K]$. Compared with the offline bound $\E[K]-\Pr[K=k]$ at the same time, the new distribution improves the gap~$c$ by at least a factor of $1+\eta$, where $\eta=2^{-k}$.

For example, start with the $(1,1)$-hard single-job distribution. Then $k=3$, and $h_1,h_2,h_3$ are chosen with probabilities $1/2,1/4,1/4$, respectively. At the earliest checkpoint~$h_3$, an offline schedule can finish all three copies and leave only~$J$, of weight~$2$. The minimum active weight at $h_1,h_2,h_3$ is thus at most $1$, $2$, and~$2$, respectively, and at most~$3/2$ in expectation. Online, the robust backlog capped at~$4$ has expectation at least $\E[K]=1+1/2+1/4=7/4$. Dividing all weights by~$3/2$ gives a gap of at least $(7/4)/(3/2)=7/6$, slightly more than the factor $1+\eta=9/8$ from above.

\paragraph{Averaging.}
Amplification improves the gap but increases the cap, making the next improvement exponentially smaller. Averaging brings the cap back down. Given a $(c,M)$-hard distribution, it concatenates~$m$ copies and divides all weights by~$m$. Their capped robust backlogs $X_i\in[0,M]$ need not be independent, but have conditional expectations at least~$c$. A second-moment bound shows that their average, capped at~$c$, has expectation at least $c-M/\sqrt m$. Thus we can reduce the cap to~$c$ while losing arbitrarily little from the gap.

\paragraph{Iterative construction.}
We fix the cap $M=\lceil4\rho\rceil$ and alternate amplification and averaging. Since the gap is at least~$1$, amplification increases it by at least~$\eta$, and averaging loses at most~$\eta/2$ while returning the cap to at most~$M$. Each round therefore adds at least~$\eta/2$ to the gap, which reaches~$2\rho$ after $2^{O(\rho)}$ rounds. Each round combines $2^{O(\rho)}$ copies, which gives the bounds on~$W$, $P$, $D$, and~$n$ in~\cref{thm:quant}.

\paragraph{Comparison with the deterministic lower bound.}
Our construction builds on the gap amplification technique of Bansal and Chan~\cite{BansalC09}, and its deterministic ingredients appear there in adaptive form. Bansal and Chan measure active weight using jobs with large remaining size, run subinstances at decreasing scales, release a final heavy job, and then also flood the instance with a stream of dense jobs. Their adversary, however, observes the algorithm. It ends each subinstance when the algorithm has fallen behind, and after the heavy job it chooses the time at which the gap is measured depending on how much of this job the algorithm has processed. An oblivious adversary can make neither choice, and three new ideas replace them.
First, hardness is a property of a distribution with a hidden terminal time, and it bounds the offline and the online active weight separately in expectation. This property survives copying, because a fresh sample remains hard for an algorithm that has adapted to all earlier samples. Second, a random checkpoint with halving probabilities replaces the adaptive choice of the time after the heavy job. No matter how the algorithm divides its time between the heavy job and the copies, the weight of the unfinished heavy job makes up in expectation for the backlogs that the algorithm clears instead. Third, this exchange is possible only if each copy contributes a bounded backlog, which is the role of the cap. Amplification increases the cap far more than the gap, so it cannot simply be repeated, and averaging restores the cap at an arbitrarily small loss in the gap.

\subsection{Further Related Work}

On a single machine, a constant competitive ratio is known for the special case of total stretch, in which the weight of every job is the reciprocal of its size. Here SRPT is $2$-competitive~\cite{MuthukrishnanRSG05}, although none of the three ratios is bounded.

For unit weights, SRPT is $O(\log\min\{P,n/m\})$-competitive on $m$ identical machines, which is best possible for $m\ge2$, even for randomized algorithms~\cite{LeonardiR07}. With weights, every randomized online algorithm on $m\ge2$ identical machines has a competitive ratio in $\Omega(\min\{\sqrt P,\sqrt W,(n/m)^{1/4}\})$~\cite{ChekuriKZ01}.

On a single machine, a constant competitive ratio becomes possible with \emph{speed augmentation}~\cite{KalyanasundaramP00,PhillipsSTW02}, where the algorithm runs on a slightly faster machine than the one used by the optimal schedule. On a machine of speed~$1+\eps$, the Highest Density First rule, which always processes a released and uncompleted job of largest density, is $O(1/\eps)$-competitive for every constant $\eps>0$~\cite{BecchettiLMP06}. With speed augmentation, constant competitive ratios for weighted flow time also extend to unrelated machines~\cite{ChadhaGKM09,AnandGK12,ChenIP25,JagerLZ26}. Other relaxations help as well. An algorithm that may reject jobs of an $O(\eps)$-fraction of the total weight can be $O(1/\eps^3)$-competitive on unrelated machines, even without preemption~\cite{LucarelliMTST18}. If the machine can vary its speed at an energy cost, then there is an $O(1)$-competitive algorithm for weighted flow time plus energy on a single machine~\cite{BansalPS09}.

Nonclairvoyant algorithms also achieve constant competitive ratios for weighted flow time with speed augmentation, both on a single machine~\cite{BansalD07} and on unrelated machines~\cite{ImKMP14}. Im, Kulkarni, and Munagala~\cite{ImKM18} introduced the more general polytope scheduling problem, in which the processing rates of the jobs are subject to arbitrary packing constraints. For weighted flow time, the nonclairvoyant Proportional Fairness algorithm is $O(1)$-competitive with $O(1)$ speed on broad subclasses of this problem~\cite{ImKM18}, whereas on general instances, every nonclairvoyant deterministic algorithm needs polylogarithmic speed to be $O(1)$-competitive~\cite{ImKM18}.
Between the clairvoyant and the nonclairvoyant setting, another line of work studies total flow time when the algorithm knows only estimates of the job sizes~\cite{AzarLT21,AzarLT22,GuptaKPW26,GuptaKLSY27}. 

\section{Preliminaries}\label{sec:prelim}

An \emph{instance}~$I$ is a finite nonempty set of jobs. We identify each job~$j$ with the triple $(r_j,p_j,w_j)$ of its release time $r_j\ge0$, size $p_j>0$, and weight $w_j>0$, so an algorithm receives no information about a job beyond these parameters. For a set~$Q$ of jobs, let $w(Q)=\sum_{j\in Q}w_j$ and $p(Q)=\sum_{j\in Q}p_j$. For~$\lambda>0$, let $I/\lambda$ and~$\lambda I$ be the instances obtained from~$I$ by dividing and multiplying all weights by~$\lambda$, respectively.

A \emph{schedule} of~$I$ assigns to each job $j\in I$ a (measurable) rate function $x_j\colon[0,\infty)\to[0,1]$ such that $x_j(t)=0$ for $t<r_j$, $\sum_{j\in I}x_j(t)\le1$ for all~$t$, and $\int_0^\infty x_j(t)\dd t\le p_j$. A schedule may thus share the machine among several jobs. This gives an online algorithm more power, and it does not change the optimum, because processing at each time an active job whose completion time in a given schedule is earliest, that is, the earliest-deadline-first rule with these completion times as deadlines, completes no job later. The \emph{remaining size} of~$j$ at time~$t$ is $p_j-\int_0^t x_j(t')\dd t'$. Job~$j$ \emph{completes} at the first time~$C_j$ at which its remaining size is zero ($C_j=\infty$ if there is no such time), and it is \emph{active} at time~$t$ if $r_j\le t<C_j$. The \emph{active weight} at time~$t$ is the total weight of the jobs that are active at time~$t$, and the weighted flow time of the schedule is its integral over time, that is,
\begin{equation*}
  \sum_{j\in I}w_j(C_j-r_j)=\int_0^\infty\sum_{j\in I\,:\,r_j\le t<C_j}w_j\dd t\,,
\end{equation*}
and $\OPT(I)$ denotes the optimum weighted flow time for instance~$I$.

A \emph{deterministic online algorithm}~$A$ maps every instance~$I$ to a schedule $A(I)$ of~$I$ subject to non-anticipation, which requires that for every $t\ge0$ and all instances $I$ and $I'$ that contain the same jobs with release times before~$t$, the schedules $A(I)$ and $A(I')$ assign the same rates to these jobs during $[0,t)$. In other words, the decisions of~$A$ before time~$t$ depend only on the jobs released before~$t$, including their sizes and weights. In particular, these jobs have the same remaining sizes at time~$t$ in $A(I)$ and in $A(I')$. We write $F_A(I)$ for the weighted flow time of $A(I)$. 

A \emph{randomized online algorithm}~$\cA$ is a probability distribution over deterministic online algorithms such that $F_A(I)$ is a random variable for every instance~$I$ when $A\sim\cA$. It is \emph{$\rho$-competitive} (against an oblivious adversary) if $\E_{A\sim\cA}[F_A(I)]\le \rho\cdot\OPT(I)$ for every instance~$I$. We use the easy direction of Yao's principle~\cite{Yao77}.

\begin{lemma}[Yao's principle]\label{lem:yao}
Let $\cD$ be a probability distribution over finitely many instances, and let $\rho>0$ be such that every deterministic online algorithm~$A$ satisfies $\E_{I\sim\cD}[F_A(I)]\ge \rho\cdot\E_{I\sim\cD}[\OPT(I)]$. Then for every randomized online algorithm~$\cA$, there is an instance~$I$ in the support of~$\cD$ with $\E_{A\sim\cA}[F_A(I)]\ge \rho\cdot\OPT(I)$.
\end{lemma}

\section{Hard Terminal Distributions}\label{sec:hard}

We first define hardness at a hidden terminal time and show how it yields a lower bound for weighted flow time. This reduces~\cref{thm:main} to the construction of hard terminal distributions in~\cref{sec:construction}.

A \emph{terminal instance} is a pair $(I,T)$ of an instance~$I$ and a \emph{terminal time} $T>0$ such that all jobs of~$I$ are released before~$T$. A \emph{terminal distribution} is a probability distribution over finitely many terminal instances. An online algorithm processes only~$I$, so its schedule $A(I)$ does not depend on~$T$. The algorithm may know the distribution in advance. Since~$I$ and~$T$ can be correlated, the jobs it sees may reveal information about~$T$.

For a terminal instance $(I,T)$, let $w^*(I,T)$ be the minimum active weight at time~$T$ over all schedules of~$I$. A job is \emph{robust} at time~$t$ in a schedule if its remaining size at~$t$ is at least~$1$. For a deterministic online algorithm~$A$, the \emph{robust backlog} $R_A(I,T)$ of~$A$ at time~$T$ is the set of jobs of~$I$ that are robust at time~$T$ in~$A(I)$. These jobs are active at time~$T$, and none of them can be completed before time~$T+1$. For $M>0$, the \emph{capped robust backlog} is
\begin{equation*}
  \varphi_A^M(I,T):=\min\bigl\{M,\,w(R_A(I,T))\bigr\}\,.
\end{equation*}

\begin{definition}\label{def:hard}
Let $0<c\le M$. A terminal distribution~$\cD$ is \emph{$(c,M)$-hard} if
\begin{enumerate}[label=(\roman*)]
	\item\label{def:hard-offline} $\E_{(I,T)\sim\cD}\bigl[w^*(I,T)\bigr]\le1$ and
	\item\label{def:hard-online} $\E_{(I,T)\sim\cD}\bigl[\varphi_A^M(I,T)\bigr]\ge c$ for every deterministic online algorithm~$A$.
\end{enumerate}
We call~$c$ the \emph{gap} and~$M$ the \emph{cap}.
\end{definition}

The cap is needed for amplification (\cref{sec:amplification}). The following observation gives the starting point of our construction and two basic properties of hardness.

\begin{observation}\label{obs:basic}
The following statements hold.
\begin{enumerate}[(a)]
\item\label{obs:base} Let~$I$ consist of a single job with release time~$0$, size~$2$, and weight~$1$, and let $T=1$. The terminal distribution supported on the single terminal instance $(I,T)$ is $(1,M)$-hard for every $M\ge1$.
\item\label{obs:monotone} If~$\cD$ is $(c,M)$-hard, then~$\cD$ is $(c',M')$-hard for all $0<c'\le c$ and $M'\ge M$.
\item\label{obs:scaling} Let $\lambda>0$ and $0<c\le M$, and let~$\cD$ be a terminal distribution such that $\E_{(I,T)\sim\cD}[w^*(I,T)]\le\lambda$ and $\E_{(I,T)\sim\cD}[\varphi_A^{\lambda M}(I,T)]\ge\lambda c$ for every deterministic online algorithm~$A$. Then the distribution of $(I/\lambda,T)$ for $(I,T)\sim\cD$ is $(c,M)$-hard.
\end{enumerate}
\end{observation}

\begin{proof}
\ref{obs:base}~No schedule completes the job by time~$1$, and in every schedule its remaining size at time~$1$ is at least~$1$. Hence $w^*(I,T)=1$ and $\varphi_A^M(I,T)=\min\{M,1\}=1$.

\ref{obs:monotone}~The offline condition in~\cref{def:hard}\ref{def:hard-offline} involves neither the gap nor the cap, so it carries over from~$(c,M)$ to~$(c',M')$. For the online condition in~\cref{def:hard}\ref{def:hard-online}, note that $M'\ge M$, so every deterministic online algorithm~$A$ and every terminal instance $(I,T)$ satisfy $\varphi_A^{M'}(I,T)\ge\varphi_A^M(I,T)$. Taking expectations gives $\E_{(I,T)\sim\cD}[\varphi_A^{M'}(I,T)]\ge\E_{(I,T)\sim\cD}[\varphi_A^M(I,T)]\ge c\ge c'$.

\ref{obs:scaling}~Since $w^*(I/\lambda,T)=w^*(I,T)/\lambda$, we have $\E_{(I,T)\sim\cD}[w^*(I/\lambda,T)]\le1$. Let~$A$ be a deterministic online algorithm. As~$A$ need not make the same scheduling decisions on~$I$ and $I/\lambda$, we apply the assumption to the algorithm~$A_\lambda$ that on input~$I$ runs~$A$ on $I/\lambda$ and uses the same rates, that is, $A_\lambda(I)=A(I/\lambda)$. Since $I$ and $I/\lambda$ have the same release times, $A_\lambda$ satisfies the non-anticipation condition. The robust backlogs of~$A$ on $(I/\lambda,T)$ and of~$A_\lambda$ on $(I,T)$ consist of the same jobs up to the scaling of their weights, so $\varphi_A^M(I/\lambda,T)=\varphi_{A_\lambda}^{\lambda M}(I,T)/\lambda$. By assumption, the right-hand side has expectation at least~$c$.
\end{proof}

\paragraph{Flooding.}
The following lemma turns a gap in active weight at the terminal time into a gap in weighted flow time by releasing a long stream of short, dense jobs. Such streams are a standard tool in lower bound constructions for flow time objectives~\cite{BecchettiLMP06,BansalC09,AzarLT22,MotwaniPT94,GuptaKLSY25,GuptaKLSY27}. In our version for terminal distributions, the stream does not depend on the algorithm.

\begin{samepage}
\begin{lemma}[Flooding]\label{lem:flooding}
Let $(I,T)$ be a terminal instance, let $M>0$, and let $L$ be a positive multiple of $d=1/(2M)$. Construct instance~$\hat I$ by adding jobs of size~$d$ and weight~$1/2$ at times $T,T+d,\dots,T+L-d$ to $I$.
\begin{enumerate}[(i)]
\item\label{it:flood-online} Every deterministic online algorithm~$A$ satisfies $F_A(\hat I)\ge L\cdot\varphi_A^M(I,T)$.
\item\label{it:flood-offline} The optimum satisfies $\OPT(\hat I)\le L\bigl(w^*(I,T)+\tfrac12\bigr)+w(I)\bigl(T+p(I)\bigr)$.
\end{enumerate}
\end{lemma}
\end{samepage}

\begin{proof}
Call the jobs of~$I$ \emph{old} and the added jobs \emph{stream jobs}.

\ref{it:flood-online}~Fix~$A$. Since the stream starts at~$T$, non-anticipation implies that each old job has the same remaining size at~$T$ in $A(I)$ and $A(\hat I)$. For $0\le\theta<L$, let~$y$ be the processing that the old jobs receive during $[T,T+\theta)$. If $y<1$, every job in $R_A(I,T)$ is still active at time~$T+\theta$, so the active weight at this time is at least~$\varphi_A^M(I,T)$. If $y\ge1$, the stream jobs released by time~$T+\theta$ have total size $d(\lfloor\theta/d\rfloor+1)>\theta$ and receive at most $\theta-y$ processing. Their total remaining size therefore exceeds~$y$. More than $y/d$ stream jobs are active, with total weight greater than $y/(2d)=My\ge M\ge\varphi_A^M(I,T)$. Thus the active weight stays at least~$\varphi_A^M(I,T)$ throughout the stream. Integration over time gives $F_A(\hat I)\ge L \cdot \varphi_A^M(I,T)$.

\ref{it:flood-offline}~Until time~$T$, follow a schedule that leaves weight $w^*(I,T)$ unfinished. Then, process every stream job immediately after its release, and after the stream finish the remaining old jobs. The old jobs contribute at most $w(I) \cdot T$ before~$T$, exactly $L\cdot w^*(I,T)$ during the stream, and at most $w(I) \cdot p(I)$ afterwards. The stream jobs contribute $L/2$. Adding these terms proves the bound.
\end{proof}

It remains to find hard terminal distributions with arbitrarily large gaps. Their existence is our main technical result, which we prove in~\cref{sec:construction}.

\begin{theorem}\label{thm:hard}
For every $\rho\ge1$, there is a $(2\rho,M)$-hard terminal distribution with $M=\lceil4\rho\rceil$.
\end{theorem}

Before that, we show that flooding such a distribution yields~\cref{thm:main}.

\begin{proof}[Proof of \cref{thm:main}]
Let $\rho\ge1$, and take a $(2\rho,M)$-hard terminal distribution~$\cD$ from~\cref{thm:hard}. Since its support is finite, there is a constant~$G$ such that $w(I)(T+p(I))\le G$ for every $(I,T)$ in the support of $\cD$. Apply~\cref{lem:flooding} with the same length~$L$ to every terminal instance in $\cD$, where $L$ is a positive multiple of $1/(2M)$ and $L\ge2G$. For the resulting distribution~$\hat\cD$ over finitely many instances, hardness and the two bounds of the lemma give
\begin{equation*}
  \E[F_A(\hat I)]-\rho\cdot \E[\OPT(\hat I)]
  \ge 2\rho L-\rho\bigl(\tfrac32L+G\bigr)
  =\rho(L/2-G)\ge0
\end{equation*}
for every deterministic online algorithm~$A$. This proves the distributional statement. For a randomized online algorithm~$\cA$, \cref{lem:yao} therefore gives an instance~$\hat I$ with $\E_{A\sim\cA}[F_A(\hat I)]\ge \rho\cdot\OPT(\hat I)$. 
\end{proof}

\section{Constructing Hard Terminal Distributions}\label{sec:construction}

We prove~\cref{thm:hard} by alternating two operations on hard terminal distributions. Amplification (\cref{sec:amplification}) multiplies the gap by $1+\eta$ for a small $\eta>0$ that depends on the cap, but it increases the cap. Averaging (\cref{sec:averaging}) reduces the cap to the input gap at an arbitrarily small loss in the gap. Both operations concatenate independent copies at decreasing time scales.

\subsection{Copying a Terminal Distribution}\label{sec:copies}

Let~$\cD$ be a terminal distribution. A \emph{bound} for~$\cD$ is a number $H\ge1$ such that $T\le H$ and $p(I)\le H$ for every terminal instance $(I,T)$ in its support. The concatenation of~$m$ copies of~$\cD$ stretches the $i$th sample in time and size by a factor~$s_i$. We define these factors backwards, together with numbers~$S_i$ that bound both the duration and the total size of the later copies $i+1,\dots,m$, and we choose~$s_i$ larger than twice~$S_i$.

\begin{definition}\label{def:concatenation}
Let~$\cD$ be a terminal distribution with a bound~$H$, and let~$m$ be a positive integer. Set $S_m=0$ and, for $i=m,\dots,1$, define $s_i=2S_i+2$ and $S_{i-1}=S_i+Hs_i$. Let $V=S_0$. Let $(I_1,T_1),\dots,(I_m,T_m)$ be independent samples from~$\cD$, and set $u_0=0$ and $u_i=u_{i-1}+s_iT_i$ for $i\in[m]$. For each job $j\in I_i$, \emph{copy~$i$} contains a job with release time $u_{i-1}+s_ir_j$, size $s_ip_j$, and weight~$w_j$. The \emph{concatenation of~$m$ copies of~$\cD$} is the union~$\bar I$ of the copies, and $\tau=u_m$ is its \emph{end}.
\end{definition}

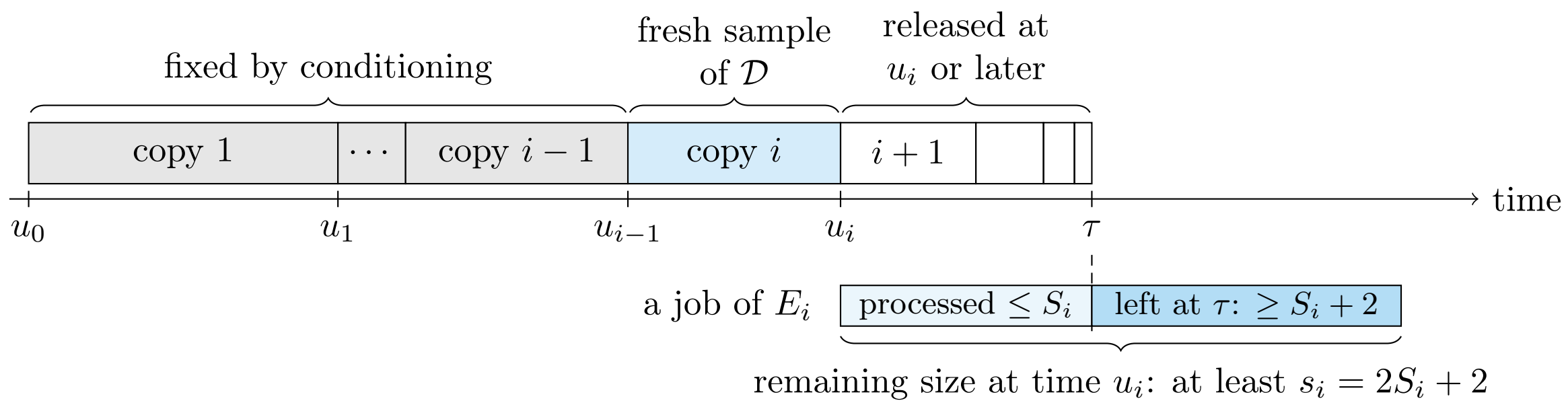
\begin{figure}[t]
  \centering
  \begin{tikzpicture}[x=0.95cm,y=1cm,font=\small]
    \def\xa{0} \def\xb{3.2} \def\xc{3.9} \def\xd{6.2} \def\xe{8.4} \def\xf{11.0} \def\xg{14.2}
    \draw[->] (-0.2,0) -- (15.0,0) node[right] {time};
    \filldraw[fill=black!10,draw=black] (\xa,0.15) rectangle (\xb,0.75);
    \node at ({(\xa+\xb)/2},0.45) {copy~$1$};
    \filldraw[fill=black!10,draw=black] (\xb,0.15) rectangle (\xc,0.75);
    \node at ({(\xb+\xc)/2},0.45) {$\cdots$};
    \filldraw[fill=black!10,draw=black] (\xc,0.15) rectangle (\xd,0.75);
    \node at ({(\xc+\xd)/2},0.45) {copy~$i-1$};
    \filldraw[fill=col1!25,draw=black] (\xd,0.15) rectangle (\xe,0.75);
    \node at ({(\xd+\xe)/2},0.45) {copy~$i$};
    \foreach \x/\y in {8.4/9.8,9.8/10.5,10.5/10.82,10.82/11.0}
      \draw (\x,0.15) rectangle (\y,0.75);
    \node at (9.1,0.45) {$i+1$};
    \draw[decorate,decoration={brace,amplitude=4pt}] (\xa+0.02,0.85) -- (\xd-0.02,0.85)
      node[midway,above=4pt] {fixed by conditioning};
    \draw[decorate,decoration={brace,amplitude=4pt}] (\xd+0.02,0.85) -- (\xe-0.02,0.85)
      node[midway,above=4pt,align=center] {fresh sample\\of~$\cD$};
    \draw[decorate,decoration={brace,amplitude=4pt}] (\xe+0.02,0.85) -- (\xf-0.02,0.85)
      node[midway,above=4pt,align=center] {released at\\$u_i$ or later};
    \foreach \x/\l in {\xa/u_0,\xb/u_1,\xd/u_{i-1},\xe/u_i,\xf/\tau} {
      \draw (\x,0.08) -- (\x,-0.08) node[below] {$\l$};
    }
    \draw[densely dashed] (\xf,-0.55) -- (\xf,-1.3);
    \filldraw[fill=col1!12,draw=black] (\xe,-0.85) rectangle (\xf,-1.25);
    \filldraw[fill=col1!45,draw=black] (\xf,-0.85) rectangle (\xg,-1.25);
    \node[font=\footnotesize] at ({(\xe+\xf)/2},-1.05) {processed $\le S_i$};
    \node[font=\footnotesize] at ({(\xf+\xg)/2},-1.05) {left at~$\tau$: $\ge S_i+2$};
    \node[anchor=east] at (\xe-0.15,-1.05) {a job of~$E_i$};
    \draw[decorate,decoration={brace,amplitude=4pt,mirror}] (\xe,-1.35) -- (\xg,-1.35)
      node[midway,below=5pt,align=center] {remaining size at time~$u_i$: at least $s_i=2S_i+2$};
  \end{tikzpicture}
  \caption{Copying a terminal distribution (\cref{lem:copies}), not to scale. The copies after copy~$i$ take at most~$S_i$ time, so every job of~$E_i$ keeps a remaining size of at least $S_i+2$ until~$\tau$.}
  \label{fig:copies}
\end{figure}

By construction, for every $i\in[m]$, we have
\begin{equation}\label{eq:scales}
  S_i=H\sum_{i'=i+1}^m s_{i'}
  \qquad\text{and}\qquad
  s_i-S_i=S_i+2\,,
\end{equation}
where the first identity also holds for~$i=0$. In particular, $V=S_0$ will bound the duration and the total size of all copies (\cref{fig:copies}).

\begin{lemma}\label{lem:copies}
Let~$\cD$ be a terminal distribution, and let~$\bar I$ be the concatenation of $m$ copies formed from independent samples $(I_1,T_1),\dots,(I_m,T_m)$ from~$\cD$, with end~$\tau$, as in~\cref{def:concatenation}.
\begin{enumerate}[(i)]
\item\label{it:geometry} The jobs of copy~$i$ are released during $[u_{i-1},u_i)$. For every $i\in\{0,\dots,m\}$, copies $i+1,\dots,m$ have total size at most~$S_i$, and $\tau-u_i\le S_i$. In particular, $p(\bar I)\le V$ and $\tau\le V$.
\item\label{it:offline} There is a schedule of~$\bar I$ in which, for every $i\in[m]$, the jobs of copy~$i$ that are active at time~$\tau$ have total weight $B_i=w^*(I_i,T_i)$. If~$\cD$ is $(c,M)$-hard, then $\E_{(I_i,T_i)\sim\cD}[B_i]\le1$.
\item\label{it:online} Suppose that~$\cD$ is $(c,M)$-hard, and let~$A$ be a deterministic online algorithm. For $i\in[m]$, let~$E_i$ be the set of jobs of copy~$i$ whose remaining size at time~$u_i$ in $A(\bar I)$ is at least~$s_i$, and let $X_i=\min\{M,w(E_i)\}$. Then $X_i$ is determined by $(I_1,T_1),\dots,(I_i,T_i)$. For every realization $(I_1,T_1),\ldots,(I_{i-1},T_{i-1})$ of the preceding $i-1$ samples of positive probability, it holds that
\begin{equation*}
  \E_{(I_i,T_i)\sim\cD}\bigl[X_i\bigm|(I_1,T_1),\dots,(I_{i-1},T_{i-1})\bigr]\ge c\,.
\end{equation*}
Moreover, every job of~$E_i$ has remaining size at least $S_i+2$ at time~$\tau$.
\end{enumerate}
\end{lemma}

\begin{proof}
\ref{it:geometry}~A job of copy~$i$ is released at time $u_{i-1}+s_ir_j$ with $0\le r_j<T_i$, hence during $[u_{i-1},u_i)$. For each $i'\in[m]$, copy~$i'$ has total size $s_{i'}\cdot p(I_{i'})\le Hs_{i'}$, and $u_{i'}-u_{i'-1}=s_{i'}T_{i'}\le Hs_{i'}$. Summing over $i'>i$ and using~\eqref{eq:scales}, we find that copies $i+1,\dots,m$ have total size at most~$S_i$ and that $\tau-u_i\le S_i$. The case $i=0$ gives $p(\bar I)\le V$ and $\tau\le V$.

\ref{it:offline}~During $[u_{i-1},u_i)$, process copy~$i$ by a schedule of~$I_i$ that leaves weight $B_i=w^*(I_i,T_i)$ unfinished at~$T_i$, stretched in time and size by~$s_i$, and do not process copy~$i$ afterwards. Stretching preserves feasibility, and the time windows are disjoint, so these schedules combine into a schedule of~$\bar I$ that leaves weight~$B_i$ of each copy~$i$ unfinished at~$\tau$. If~$\cD$ is $(c,M)$-hard, then $(I_i,T_i)$ is a sample from a hard distribution, so $\E_{(I_i,T_i)\sim\cD}[B_i]=\E_{(I_i,T_i)\sim\cD}[w^*(I_i,T_i)]\le1$ by~\cref{def:hard}\ref{def:hard-offline}.

\ref{it:online}~Fix $i\in[m]$ and a realization of the first $i-1$ samples $(I_1,T_1),\dots,(I_{i-1},T_{i-1})$ of positive probability. This fixes copies $1,\dots,i-1$, their union~$\bar I_{<i}$, and the time~$u_{i-1}$. With these copies fixed, the behavior of~$A$ on copy~$i$ defines a deterministic online algorithm~$A_i$ for the unstretched sample~$I_i$. Since~$(I_i,T_i)$ is a fresh sample from~$\cD$, the hardness of~$\cD$ applies to~$A_i$.

To define~$A_i$, let $\sigma_i(j)=(u_{i-1}+s_ir_j,\,s_ip_j,\,w_j)$ for a job $j=(r_j,p_j,w_j)$, and let $\sigma_i(I)=\{\sigma_i(j):j\in I\}$ for an instance~$I$. Thus copy~$i$ is~$\sigma_i(I_i)$. On input~$I$, algorithm~$A_i$ runs~$A$ on $\bar I_{<i}\cup\sigma_i(I)$ and, at time~$t$, gives each job $j\in I$ the rate that~$A$ gives~$\sigma_i(j)$ at time $u_{i-1}+s_it$. Processing that~$A$ assigns to~$\bar I_{<i}$ becomes idle time for~$A_i$. Since time and sizes are stretched by the same factor~$s_i$, these rates form a schedule of~$I$, and the remaining size of~$j$ at time~$t$ in $A_i(I)$ is the remaining size of~$\sigma_i(j)$ at time $u_{i-1}+s_it$ in $A(\bar I_{<i}\cup\sigma_i(I))$, divided by~$s_i$. Moreover, $A_i$ is a deterministic online algorithm. If two inputs contain the same jobs with release times before~$t$, then the instances on which~$A$ runs contain the same jobs with release times before $u_{i-1}+s_it$, so~$A_i$ inherits non-anticipation from~$A$.

Now consider the input~$I_i$. The instance $\bar I_{<i}\cup\sigma_i(I_i)$ consists of copies $1,\dots,i$. It contains the same jobs as~$\bar I$ with release times before~$u_i$, because copies $i+1,\dots,m$ are released at~$u_i$ or later by~\ref{it:geometry}. By non-anticipation, the jobs of copy~$i$ therefore have the same remaining sizes at time~$u_i$ in $A(\bar I)$ and in $A(\bar I_{<i}\cup\sigma_i(I_i))$. Since $u_i=u_{i-1}+s_iT_i$, a job~$\sigma_i(j)$ of copy~$i$ belongs to~$E_i$ if and only if~$j$ has remaining size at least~$1$ at time~$T_i$ in $A_i(I_i)$, that is, if and only if $j\in R_{A_i}(I_i,T_i)$. As~$\sigma_i$ preserves weights, we obtain $X_i=\varphi_{A_i}^M(I_i,T_i)$. Since~$A_i$ depends only on the first $i-1$ samples, $X_i$ is determined by the first~$i$ samples.

Since~$A_i$ does not depend on~$(I_i,T_i)$ and the samples are independent, $X_i$ is the capped robust backlog of a fixed algorithm on a fresh sample from~$\cD$. As~$\cD$ is $(c,M)$-hard, applying~\cref{def:hard}\ref{def:hard-online} to~$A_i$ yields
\begin{equation*}
  \E_{(I_i,T_i)\sim\cD}\bigl[X_i\bigm|(I_1,T_1),\dots,(I_{i-1},T_{i-1})\bigr]
  =\E_{(I,T)\sim\cD}\bigl[\varphi_{A_i}^M(I,T)\bigr]
  \ge c\,.
\end{equation*}

Finally, a job of~$E_i$ has remaining size at least~$s_i$ at~$u_i$ and receives at most $\tau-u_i\le S_i$ processing before~$\tau$. Its remaining size at~$\tau$ is therefore at least $s_i-S_i=S_i+2$.
\end{proof}

\subsection{Amplification}\label{sec:amplification}

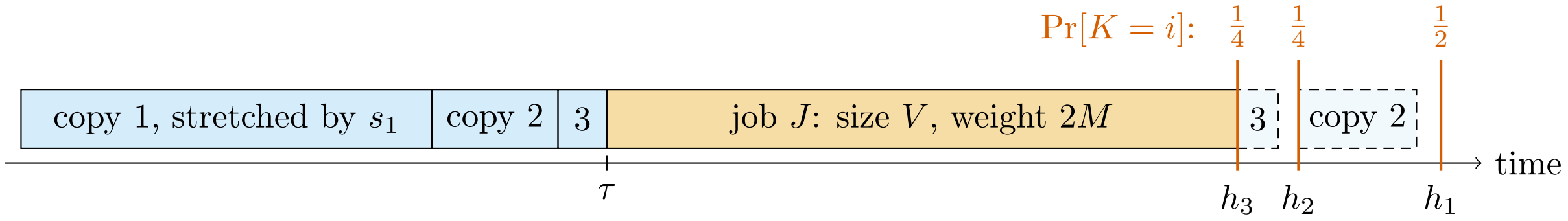
\begin{figure}[t]
  \centering
  \begin{tikzpicture}[x=0.83cm,y=1cm,font=\small]
    \def\ub{5.05} \def\uc{6.6} \def\ud{7.2}
    \def\hc{14.95} \def\hb{15.7} \def\ha{17.45}
    \draw[->] (-0.2,0) -- (17.95,0) node[right] {time};
    \foreach \x/\y/\l in {0/\ub/{copy~$1$, stretched by $s_1$},\ub/\uc/{copy~$2$},\uc/\ud/{$3$}} {
      \filldraw[fill=col1!25,draw=black] (\x,0.15) rectangle (\y,0.75);
      \node at ({(\x+\y)/2},0.45) {\l};
    }
    \filldraw[fill=col2!35,draw=black] (\ud,0.15) rectangle (\hc,0.75);
    \node at ({(\ud+\hc)/2},0.45) {job $J$: size $V$, weight $2M$};
    \foreach \x/\w/\l in {\hc/0.5/{$3$},\hb/1.45/{copy~$2$}} {
      \draw[densely dashed,fill=col1!8] (\x,0.15) rectangle (\x+\w,0.75);
      \node at (\x+\w/2,0.45) {\l};
    }
    \draw (\ud,0.08) -- (\ud,-0.08) node[below] {$\tau$};
    \foreach \x/\l/\p in {\hc/h_3/{\tfrac14},\hb/h_2/{\tfrac14},\ha/h_1/{\tfrac12}} {
      \draw[col5,thick] (\x,1.05) -- (\x,-0.08) node[below,black] {$\l$};
      \node[above,col5] at (\x,1.05) {$\p$};
    }
    \node[anchor=south east,col5] at (\hc-0.35,1.05) {$\Pr[K=i]$:};
  \end{tikzpicture}
  \caption{Amplification for $M=1$ and $k=3$, before weight normalization (not to scale). By the checkpoint~$h_i$, a schedule can complete~$J$ and then the remaining jobs of copies $i+1,\dots,k$ (dashed). The terminal time is~$h_K$, where $K=i$ with the probability shown above~$h_i$.}
  \label{fig:amplification}
\end{figure}

Amplification appends a heavy job to $k$ copies of a terminal distribution and hides the terminal time among $k$ checkpoints after the release of this job (\cref{fig:amplification}). Let $M\ge1$ be an integer and $k=2M+1$. Let~$\cD$ be a terminal distribution with a bound~$H$, and let~$\bar I$ be the concatenation of~$k$ copies of~$\cD$ (\cref{def:concatenation}), with the notation of~\cref{sec:copies} for $m=k$. We add a job~$J$ with release time~$\tau$, size~$V$, and weight~$2M$, and we define for each $i\in[k]$ the \emph{checkpoint} $h_i=\tau+V+S_i$. Since $S_1>S_2>\dots>S_k=0$, we have $h_1>h_2>\dots>h_k=\tau+V$. By~\cref{lem:copies}\ref{it:geometry}, copies $i+1,\dots,k$ have total size at most~$S_i$, and all copies together have total size at most~$V$. Hence the time $V+S_i$ from~$\tau$ to~$h_i$ suffices to complete either~$J$ and copies $i+1,\dots,k$, or all copies.

The terminal time for the new distribution is the checkpoint~$h_K$ for a random index $K\in[k]$, drawn independently of the copies with $\Pr[K\ge i]=2^{1-i}$ for all $i\in[k]$. Equivalently, $K$ is the number of fair coin flips up to and including the first head, truncated at~$k$. Thus, for every $b\in\{0,\dots,k\}$,
\begin{equation}\label{eq:tail}
  \sum_{\ell=b+1}^k\Pr[K\ge \ell]\le2\Pr[K>b]\,,
\end{equation}
because the left-hand side equals $2^{1-b}-2^{1-k}$ and $\Pr[K>b]=2^{-b}$ for $b<k$, and both sides vanish for $b=k$. Moreover, $\E[K]=\sum_{i=1}^k\Pr[K\ge i]=2-2^{1-k}$ and $\Pr[K=k]=2^{1-k}$. Let $\beta :=\E[K]-\Pr[K=k]=2-2^{2-k}$. The \emph{amplified} terminal distribution $\Amp_M(\cD)$ is the distribution of $\bigl((\bar I\cup\{J\})/\beta,h_K\bigr)$.

\begin{lemma}\label{lem:amplification}
Let $M\ge1$ be an integer, $k=2M+1$, and $\eta=2^{-k}$. If~$\cD$ is $(c,M)$-hard, then $\Amp_M(\cD)$ is $((1+\eta)c,(k+1)M)$-hard. Moreover, every terminal instance $(I,T)$ in its support satisfies $w^*(I,T)\le2M$.
\end{lemma}

\begin{proof}
All jobs of $\bar I\cup\{J\}$ are released by time $\tau<h_K$, and the support is finite, so $\Amp_M(\cD)$ is a terminal distribution. We show that, before the weights are divided by~$\beta$,
\begin{subequations}\label{eq:amp-goal}
\begin{align}
  \E\bigl[w^*(\bar I\cup\{J\},h_K)\bigr]&\le\beta\,,\label{eq:amp-goal-offline}\\*
  \E\bigl[\varphi_A^{(k+1)M}(\bar I\cup\{J\},h_K)\bigr]&\ge c\cdot \E[K]\,,\label{eq:amp-goal-online}
\end{align}
\end{subequations}
for every deterministic online algorithm~$A$, where the expectations are over the copies and~$K$.

We first derive the lemma from~\eqref{eq:amp-goal}. Since $k\ge3$, we have $1\le\beta\le2$. As $\E[K]=\beta+2^{1-k}$, it follows that
\begin{equation*}
  \frac{\E[K]}{\beta}=1+\frac{2^{1-k}}{\beta}\ge1+2^{-k}=1+\eta\,.
\end{equation*}
Raising the cap in~\eqref{eq:amp-goal-online} from $(k+1)M$ to $\beta(k+1)M$ can only increase the capped robust backlog, so
\begin{equation*}
  \E\bigl[\varphi_A^{\beta(k+1)M}(\bar I\cup\{J\},h_K)\bigr]\ge c\cdot \E[K]\ge\beta(1+\eta)c\,.
\end{equation*}
With~\eqref{eq:amp-goal-offline}, this verifies the assumptions of~\cref{obs:basic}\ref{obs:scaling} for the distribution of $(\bar I\cup\{J\},h_K)$, with $\lambda=\beta$, gap $(1+\eta)c$, and cap $(k+1)M$. Here $(1+\eta)c\le2M\le(k+1)M$ because $c\le M$. Hence $\Amp_M(\cD)$, the distribution of $\bigl((\bar I\cup\{J\})/\beta,h_K\bigr)$, is $((1+\eta)c,(k+1)M)$-hard.

\emph{Proof of~\eqref{eq:amp-goal-offline}.} Follow the schedule of~\cref{lem:copies}\ref{it:offline} until~$\tau$, which leaves weight~$B_\ell$ of copy~$\ell$ unfinished at~$\tau$, where $\E[B_\ell]\le1$. By the choice of~$h_i$, it can then complete during $[\tau,h_i)$ either~$J$ and the remaining jobs of copies $i+1,\dots,k$, or all remaining jobs of~$\bar I$. Thus
\begin{equation*}
  w^*(\bar I\cup\{J\},h_i)\le\min\left\{\sum_{\ell\le i}B_\ell,\,2M\right\}\,.
\end{equation*}
In particular, $w^*(\bar I\cup\{J\},h_K)\le2M$ for every realization of the copies and~$K$. Dividing the weights by~$\beta\ge1$ gives $w^*(I,T)\le2M$ for every terminal instance $(I,T)$ in the support of~$\Amp_M(\cD)$, which is the second claim of the lemma.

To bound the expectation, fix $i\in[k]$. Since~$K$ is independent of the copies, $\E[B_\ell\mid K=i]=\E[B_\ell]\le1$ for every~$\ell$. As the minimum is at most each of its arguments, taking conditional expectations yields
\begin{equation*}
  \E\bigl[w^*(\bar I\cup\{J\},h_K)\bigm|K=i\bigr]\le\min\left\{\sum_{\ell\le i}\E[B_\ell\mid K=i],\,2M\right\}\le\min\{i,2M\}\,.
\end{equation*}
Averaging over~$K$ and using $2M=k-1$ gives
\begin{equation*}
  \E\bigl[w^*(\bar I\cup\{J\},h_K)\bigr]\le\E[\min\{K,k-1\}]=\E[K]-\Pr[K=k]=\beta\,.
\end{equation*}

\emph{Proof of~\eqref{eq:amp-goal-online}.} Fix a deterministic online algorithm~$A$. Let $Z=((I_1,T_1),\dots,(I_k,T_k))\sim\cD^k$ denote the tuple of independent samples that form the copies, and fix a realization of~$Z$. The variable~$K$ affects only the terminal time, so the schedule $A(\bar I\cup\{J\})$ is fixed, and only the checkpoint at which we evaluate it is random. Let~$E_\ell$ and~$X_\ell$ be as in~\cref{lem:copies}\ref{it:online}. Since~$J$ is released at time~$\tau$, non-anticipation implies that $A(\bar I\cup\{J\})$ and $A(\bar I)$ process the jobs of~$\bar I$ identically before~$\tau$. Hence every job of~$E_\ell$ has remaining size at least $S_\ell+2$ at~$\tau$ also in $A(\bar I\cup\{J\})$.

Let $b\in\{0,\dots,k\}$ be the largest index~$\ell$ such that~$J$ is not robust at~$h_\ell$, with $b=0$ if~$J$ is robust at every checkpoint. The remaining size of~$J$ does not increase over time and $h_1>\dots>h_k$, so~$J$ is robust at~$h_i$ if and only if $i>b$.

We claim that for every $\ell\le b$, all jobs of~$E_\ell$ are robust at every checkpoint~$h_i$ with $i\ge \ell$. Indeed, $J$ is not robust at~$h_\ell$, so it received more than $V-1$ processing during $[\tau,h_\ell)$, which leaves less than $(V+S_\ell)-(V-1)=S_\ell+1$ processing for the jobs of~$\bar I$. Every job of~$E_\ell$ therefore has remaining size greater than~$1$ at~$h_\ell$, and hence also at every earlier checkpoint~$h_i$ with $i\ge \ell$.

Now fix a checkpoint~$h_i$. By the claim, its robust backlog contains the sets~$E_\ell$ with $\ell\le\min\{i,b\}$, and it contains~$J$ if $i>b$. The sets $E_1,\dots,E_k$ and~$\{J\}$ are pairwise disjoint, and $X_\ell\le w(E_\ell)$, so the weight of the robust backlog at~$h_i$ is at least
\begin{equation*}
  Y_i:=\sum_{\ell\le\min\{i,b\}}X_\ell+2M\cdot\bone_{\{i>b\}}\,.
\end{equation*}
We also have $Y_i\le(k+1)M$. Indeed, since $X_\ell\le M$, the sum~$Y_i$ is at most~$kM$ if $i\le b$, and at most $(k-1)M+2M$ if $i>b$, because then $b\le k-1$. The capped robust backlog $\varphi_A^{(k+1)M}(\bar I\cup\{J\},h_i)$ is the minimum of $(k+1)M$ and the weight of the robust backlog at~$h_i$, and therefore at least~$Y_i$.

We now take expectations, first over~$K$ with~$Z$ fixed. The copies, the schedule, the index~$b$, and the values~$X_\ell$ remain fixed, and the distribution of~$K$ is unchanged because~$K$ is independent of~$Z$. In~$Y_K$, a copy $\ell\le b$ contributes~$X_\ell$ whenever $K\ge \ell$, and~$J$ contributes~$2M$ whenever $K>b$, so
\begin{align*}
  \E_K\bigl[\varphi_A^{(k+1)M}(\bar I\cup\{J\},h_K)\bigm|Z\bigr]
  \ge\E_K\bigl[Y_K\bigm|Z\bigr]
  &=\sum_{\ell\le b}\Pr[K\ge \ell]\cdot X_\ell+2M \cdot \Pr[K>b]\\*
  &\ge\sum_{\ell=1}^k\Pr[K\ge \ell]\cdot X_\ell\,,
\end{align*}
where the final inequality uses \eqref{eq:tail} and for every $\ell > b$ that $X_\ell \le M$.

We now average over~$Z\sim\cD^k$. Since~$X_\ell$ is determined by the first~$\ell$ samples, averaging the conditional bound of~\cref{lem:copies}\ref{it:online} over the first $\ell-1$ samples gives $\E_{Z\sim\cD^k}[X_\ell]\ge c$ for every~$\ell$. The law of total expectation therefore yields
\begin{align*}
  \E_{Z,K}\bigl[\varphi_A^{(k+1)M}(\bar I\cup\{J\},h_K)\bigr]
  &=\E_{Z\sim\cD^k}\Bigl[\E_K\bigl[\varphi_A^{(k+1)M}(\bar I\cup\{J\},h_K)\bigm|Z\bigr]\Bigr]\\
  &\ge\sum_{\ell=1}^k\Pr[K\ge \ell]\cdot \E_{Z\sim\cD^k}[X_\ell]\\
  &\ge c\sum_{\ell=1}^k\Pr[K\ge \ell]=c\cdot \E_K[K]\,.
\end{align*}
This is~\eqref{eq:amp-goal-online} and thus completes the proof.
\end{proof}

\subsection{Averaging}\label{sec:averaging}

For a terminal distribution~$\cD$ with a bound~$H$ and a positive integer~$m$, the \emph{averaged} terminal distribution $\Avg_m(\cD)$ is the distribution of the terminal instance $(\bar I/m,\tau)$, where~$\bar I$ is the concatenation of~$m$ copies of~$\cD$ (\cref{def:concatenation}), formed with the bound~$H$, and~$\tau$ is its end.

\begin{lemma}\label{lem:averaging}
Let~$\cD$ be a $(c,M)$-hard terminal distribution and let~$m$ be a positive integer. For $(I,T)\sim\Avg_m(\cD)$, we have $\E[w^*(I,T)]\le1$ and $\E[\varphi_A^c(I,T)]\ge c-\frac{M}{\sqrt m}$
for every deterministic online algorithm~$A$.
In particular, if $c>M/\sqrt m$, then $\Avg_m(\cD)$ is $(c-M/\sqrt m,c)$-hard.
\end{lemma}

\begin{proof}
All jobs of~$\bar I$ are released before~$\tau$ by~\cref{lem:copies}\ref{it:geometry}, so $\Avg_m(\cD)$ is a terminal distribution. For every instance $I$, we have $w^*(I/m,T)=w^*(I,T)/m$. Moreover, for a deterministic online algorithm~$A$, the algorithm~$A'$ with $A'(I)=A(I/m)$ is a deterministic online algorithm with $\varphi_{A'}^{cm}(\bar I,\tau)=m\cdot\varphi_A^c(\bar I/m,\tau)$, as in the proof of~\cref{obs:basic}\ref{obs:scaling}. It therefore suffices to show that
\begin{equation*}
  \E[w^*(\bar I,\tau)]\le m
  \qquad\text{and}\qquad
  \E[\varphi_A^{cm}(\bar I,\tau)]\ge cm-M\sqrt m
\end{equation*}
for every deterministic online algorithm~$A$. The first bound holds because \cref{lem:copies}\ref{it:offline} gives $w^*(\bar I,\tau)\le B_1+\dots+B_m$ with $\E[B_i]\le1$.

To show the second bound, fix a deterministic online algorithm~$A$, and let~$E_i$ and~$X_i$ be as in~\cref{lem:copies}\ref{it:online}. Every job of~$E_i$ has remaining size at least $S_i+2\ge2$ at~$\tau$, so the robust backlog at~$\tau$ contains the pairwise disjoint sets $E_1,\dots,E_m$. Since $X_i\le w(E_i)$, we obtain
\begin{equation*}
  \varphi_A^{cm}(\bar I,\tau)\ge\min\left\{cm,\sum_{i=1}^mX_i\right\}=cm-\Delta_m\,,
\end{equation*}
where $\Delta_m=\bigl(cm-\sum_{i=1}^mX_i\bigr)^+$. It remains to show that $\E[\Delta_m]\le M\sqrt m$.

We bound~$\E[\Delta_m^2]$ by induction over the copies. For $i=0,\dots,m$, let $\Delta_i=\bigl(ic-\sum_{\ell=1}^iX_\ell\bigr)^+$. We show that $\E[\Delta_i^2]\le iM^2$ for all~$i$. The case $i=0$ holds because $\Delta_0=0$. Now let $i\in[m]$. Since $(i-1)c-\sum_{\ell<i}X_\ell\le \Delta_{i-1}$ and $x\mapsto x^+$ is nondecreasing, we have $\Delta_i\le(\Delta_{i-1}+c-X_i)^+$. Squaring and using $0\le X_i\le M$ and $c\le M$, we obtain
\begin{equation*}
  \Delta_i^2\le(\Delta_{i-1}+c-X_i)^2\le \Delta_{i-1}^2+2\Delta_{i-1}(c-X_i)+M^2\,.
\end{equation*}
Fix a realization of the first $i-1$ samples of positive probability. By~\cref{lem:copies}\ref{it:online}, these samples determine $X_1,\dots,X_{i-1}$ and hence the value $\Delta_{i-1}\ge0$. By the same lemma, the conditional expectation of~$X_i$ given these samples is at least~$c$. Therefore
\begin{equation*}
  \E\bigl[2\Delta_{i-1}(c-X_i)\bigm|(I_1,T_1),\dots,(I_{i-1},T_{i-1})\bigr]
  =2\Delta_{i-1}\Bigl(c-\E\bigl[X_i\bigm|(I_1,T_1),\dots,(I_{i-1},T_{i-1})\bigr]\Bigr)\le0\,.
\end{equation*}
By the law of total expectation, the term $2\Delta_{i-1}(c-X_i)$ has nonpositive expectation. Thus, taking expectations in the bound on~$\Delta_i^2$ gives $\E[\Delta_i^2]\le\E[\Delta_{i-1}^2]+M^2\le iM^2$ by the induction hypothesis. Finally, for $i=m$, Cauchy--Schwarz gives $\E[\Delta_m]\le\sqrt{\E[\Delta_m^2]}\le M\sqrt m$.
\end{proof}

\subsection{Iterative Construction of a Hard Terminal Distribution}\label{sec:iteration}

It remains to combine the two operations.

\begin{proof}[Proof of~\cref{thm:hard}]
Fix $\rho\ge1$ and let $M=\lceil4\rho\rceil$. This will be the cap of the distributions~$\cD_q$ constructed below, and the last of them will have gap $M/2\ge2\rho$. Set
\begin{equation*}
  k=2M+1,\qquad \eta=2^{-k},\qquad
  m=\left(\frac{2(k+1)M}{\eta}\right)^2,\qquad
  N=\frac{M-2}{\eta}\,.
\end{equation*}
Here~$k$ and~$\eta$ are the parameters of~\cref{lem:amplification}, and~$m$ is chosen so that averaging an input with cap $(k+1)M$ loses at most $(k+1)M/\sqrt m=\eta/2$ by~\cref{lem:averaging}. Since $M\ge4$ and $1/\eta=2^k$, both~$m$ and~$N$ are positive integers. Let~$\cD_0$ be the single-job distribution of~\cref{obs:basic}\ref{obs:base}, and for $q=1,\dots,N$ define iteratively the distribution
\begin{equation*}
  \cD_q=\Avg_m(\Amp_M(\cD_{q-1}))\,.
\end{equation*}
We call one amplification followed by one averaging a \emph{round}, so~$N$ is the number of rounds. Both operations depend on a bound for their input distribution through the scales of~\cref{def:concatenation}. \Cref{lem:amplification,lem:averaging} hold for any bound, and we fix specific bounds in~\cref{sec:parameters}.

We show by induction on~$q$ that~$\cD_q$ is $(c_q,M)$-hard, where $c_q=1+q\eta/2$. For $q=0$, this is~\cref{obs:basic}\ref{obs:base}. Let $q\in[N]$, and suppose that~$\cD_{q-1}$ is $(c_{q-1},M)$-hard. By~\cref{lem:amplification}, amplification gives gap $(1+\eta)c_{q-1}$ and cap $(k+1)M$. By~\cref{lem:averaging}, averaging then loses at most $(k+1)M/\sqrt m=\eta/2$ from the gap and reduces the cap to $(1+\eta)c_{q-1}$. Since $c_{q-1}\ge1$, the gap of~$\cD_q$ is at least
\begin{equation*}
  (1+\eta)c_{q-1}-\eta/2\ge c_{q-1}+\eta/2=c_q\,.
\end{equation*}
Its cap satisfies $(1+\eta)c_{q-1}<M$, because $c_{q-1}<c_N=M/2$ and $\eta<1$. By~\cref{obs:basic}\ref{obs:monotone}, $\cD_q$ is $(c_q,M)$-hard. Finally, we have $c_N=M/2\ge2\rho$, which proves the theorem.
\end{proof}

\section{Quantitative Bounds and Extensions}\label{sec:quantitative}

We bound the parameters of our instances and prove~\cref{thm:quant} in~\cref{sec:parameters}. We then prove the high-probability bound of~\cref{thm:probability} in~\cref{sec:probability} and the bound for nearly known instances of~\cref{thm:interarrival} in~\cref{sec:interarrival}.

\subsection{Parameter Bounds}\label{sec:parameters}

For an instance~$I$, let $W(I)=\max_{j\in I}w_j/\min_{j\in I}w_j$ and $P(I)=\max_{j\in I}p_j/\min_{j\in I}p_j$ be its weight and size ratios. Let $D(I)=\max_{j\in I}(w_j/p_j)/\min_{j\in I}(w_j/p_j)$ be its density ratio, and let~$|I|$ be its number of jobs. Bounding these parameters in the construction of~\cref{sec:iteration} gives the following refinement of~\cref{thm:main}.

\begin{theorem}\label{thm:quant-refined}
There is a constant $C\ge1$ such that for every $\rho\ge1$, there is a probability distribution~$\hat\cD$ over finitely many instances such that every instance~$\hat I$ in its support satisfies
\begin{equation*}
  W(\hat I)\le2^{2^{C\rho}}\qquad \text{and} \qquad \max\{P(\hat I),D(\hat I),|\hat I|\}\le2^{2^{2^{C\rho}}}\,,
\end{equation*}
and every deterministic online algorithm~$A$ satisfies $\E_{\hat I\sim\hat\cD}[F_A(\hat I)]\ge \rho\cdot\E_{\hat I\sim\hat\cD}[\OPT(\hat I)]$.
\end{theorem}

\begin{proof}
Let $M$, $k$, $\eta$, $m$, $N$, and $\cD_0,\dots,\cD_N$ be as in~\cref{sec:iteration}. We obtain~$\hat\cD$ by flooding~$\cD_N$ (\cref{lem:flooding}) with a common stream length~$L$ chosen below. The stream jobs have weight~$1/2$ and size $1/(2M)$.

\emph{Weights.} All weights in~$\cD_q$ lie in $[(2m)^{-q},1]$. Indeed, the initial weight is~$1$. Each round adds a heavy job of weight~$2M$ and divides all weights by~$\beta m\in[m,2m]$, where $\beta\in[1,2]$ is the normalization of amplification. The claim follows by induction, since $m\ge2M$. The stream weight~$1/2$ lies in the same interval for $q=N\ge1$, so $W(\hat I)\le(2m)^N$.

\emph{Sizes.} The backward recurrence for the copy scales in~\cref{def:concatenation} gives $S_{i-1}+1=(2H+1)(S_i+1)$. Since $S_r=0$ for a concatenation of~$r$ copies, and $2H+1\le H^2$ for $H\ge3$, such a concatenation with bound $H\ge3$ has $V=(2H+1)^r-1\le H^{2r}$.
The output of amplification with bound~$H$ has total size at most~$2V$ and terminal times at most $\tau+V+S_1\le3V$. Thus $H^{3k}$ is a bound for it, since $3V\le3H^{2k}\le H^{3k}$. Averaging with the bound~$H^{3k}$ then yields total sizes and terminal times at most $(H^{3k})^{2m}=H^{6km}$, by~\cref{lem:copies}\ref{it:geometry}. We therefore fix the bounds as follows. In round~$q$, amplification uses the bound~$H_{q-1}$, and averaging uses the bound~$H_{q-1}^{3k}$, where $H_0=3$ and $H_q=H_{q-1}^{6km}=3^{(6km)^q}$. By induction, $H_q$ is a bound for~$\cD_q$.

All jobs in~$\cD_q$ also have size at least~$2$, because the initial job has size~$2$, copying stretches sizes by factors $s_i\ge2$, and every heavy job has size $V\ge2H\ge2$. After flooding, all sizes lie in $[1/(2M),H_N]$, so $P(\hat I)\le2MH_N$.

\emph{Densities.} For every instance~$I$, the ratio of any two job densities satisfies $(w_j/p_j)/(w_{j'}/p_{j'})=(w_j/w_{j'})\cdot(p_{j'}/p_j)\le W(I)\cdot P(I)$. Hence $D(I)\le W(I)P(I)$, and the preceding bounds give $D(\hat I)\le2MH_N(2m)^N$.

\emph{Cost and number of jobs.} An instance~$I$ in the support of~$\cD_N$ consists of jobs of size at least~$2$ and has total size at most~$H_N$, so it has at most $H_N/2$ jobs, each of weight at most~$1$. Hence $w(I)\le H_N/2$, and the additive term $w(I)(T+p(I))$ in~\cref{lem:flooding} is at most $H_N^2$. Choose $L=2H_N^2+2$, a multiple of $1/(2M)$ because~$H_N$ and~$M$ are integers. Since~$\cD_N$ is $(2\rho,M)$-hard, the lemma gives
\begin{equation*}
  \E[F_A(\hat I)]-\rho\cdot \E[\OPT(\hat I)]
  \ge2\rho L-\rho\bigl(\tfrac32L+H_N^2\bigr)
  =\rho(L/2-H_N^2)=\rho>0
\end{equation*}
for every deterministic online algorithm~$A$, as required. The stream adds $2ML=2M(2H_N^2+2)$ jobs. As $M\ge4$ and $H_N\ge3$, we obtain $|\hat I|\le H_N/2+2M(2H_N^2+2)\le(2MH_N)^2$.

We have $M\le5\rho$ and $k\le11\rho$. The explicit parameters of the iteration give
\begin{equation*}
  \log m=2k+2\log(2(k+1)M)=O(\rho) \qquad \text{and} \qquad
  N=(M-2)2^k=2^{O(\rho)}\,.
\end{equation*}
Consequently,
\begin{equation*}
  \log W(\hat I)\le N\log(2m)=2^{O(\rho)} \qquad \text{and} \qquad
  \log\log H_N=N\log(6km)+O(1)=2^{O(\rho)}\,.
\end{equation*}
Together with $P(\hat I)\le2MH_N$, $D(\hat I)\le2MH_N(2m)^N$, and $|\hat I|\le(2MH_N)^2$, these estimates give the claimed bounds with an absolute constant~$C$.
\end{proof}

\begin{proof}[Proof of \cref{thm:quant}]
Let $W,P,D,n\ge16$, and let~$C$ be the constant of~\cref{thm:quant-refined}. Fix a randomized online algorithm~$\cA$ and set $\rho=\min\{\log\log W,\log\log\log\min\{P,D,n\}\}/C$. This is positive because $W,P,D,n\ge16$. If $\rho<1$, there is nothing to show.
Otherwise, let~$\hat\cD$ be the distribution of~\cref{thm:quant-refined}. By~\cref{lem:yao}, there is an instance~$I$ in its support with
\begin{equation*}
  W(I)\le2^{2^{C\rho}}\le W \qquad \text{and} \qquad
  \max\{P(I),D(I),|I|\}\le2^{2^{2^{C\rho}}}\le\min\{P,D,n\}
\end{equation*}
and $\E_{A\sim\cA}[F_A(I)]\ge \rho\cdot\OPT(I)$. Since $\OPT(I)>0$, the algorithm~$\cA$ is not $\rho'$-competitive on this class of instances for any $\rho'<\rho$, which proves the claimed lower bound.
\end{proof}

\subsection{High-Probability Bounds}\label{sec:probability}

The last averaging step also yields the high-probability lower bound of~\cref{thm:probability}.

\begin{proof}[Proof of \cref{thm:probability}]
Use the construction of~\cref{sec:iteration} with~$2\rho$ in place of~$\rho$. Thus $M=\lceil8\rho\rceil$, $k=2M+1$, $\eta=2^{-k}$, and $m=(2(k+1)M/\eta)^2$, and~$\cD_N$ is obtained by averaging $\Amp_M(\cD_{N-1})$. This amplified distribution has gap $(1+\eta)c_{N-1}\ge c_N=M/2$ and cap $(k+1)M$, so it is $(M/2,(k+1)M)$-hard by~\cref{obs:basic}\ref{obs:monotone}. A sample of~$\cD_N$ is $(I,T)=(\bar I/m,\tau)$, where~$\bar I$ is the concatenation of~$m$ copies of the amplified distribution and~$\tau$ is its end. Averaging the capped robust backlogs of the copies and capping the result at~$M$ gives a lower bound on~$\varphi_A^M(I,T)$ for every deterministic online algorithm~$A$. The average of their minimum active weights gives an upper bound on~$w^*(I,T)$.

Fix a deterministic online algorithm~$A$, and let~$A'$ be the algorithm with $A'(I')=A(I'/m)$ for every instance~$I'$, as in the proof of~\cref{lem:averaging}. Let~$E_i$ and~$X_i$ be as in~\cref{lem:copies}\ref{it:online} for the algorithm~$A'$, the amplified distribution, and its cap $(k+1)M$. As in the proof of~\cref{lem:averaging}, the robust backlog of~$A'$ at~$\tau$ contains the pairwise disjoint sets $E_1,\dots,E_m$, and the robust backlog of~$A$ on $(\bar I/m,\tau)$ consists of the same jobs with weights divided by~$m$. Hence
\begin{equation*}
  \varphi_A^M(I,T)\ge\min\biggl\{M,\,\frac1m\sum_{i=1}^mX_i\biggr\}\,.
\end{equation*}
By~\cref{lem:copies}\ref{it:online}, each $X_i\in[0,(k+1)M]$ is determined by the first~$i$ samples, and its conditional expectation~$\mu_i$ given the first $i-1$ samples is at least $M/2\ge4\rho$. The differences $X_i-\mu_i$ therefore form a martingale difference sequence with $\abs{X_i-\mu_i}\le(k+1)M$, and the Azuma--Hoeffding inequality~\cite{Azuma67,Hoeffding63} gives
\begin{equation*}
  \Pr\biggl[\sum_{i=1}^mX_i<3\rho m\biggr]
  \le\Pr\biggl[\sum_{i=1}^m(X_i-\mu_i)<-\rho m\biggr]
  \le\exp\biggl(-\frac{\rho^2m}{2(k+1)^2M^2}\biggr)
  =\exp\bigl(-2\rho^2\cdot4^k\bigr)\,,
\end{equation*}
where the last step uses $m=4(k+1)^2M^2\cdot4^k$. Since $M\ge3\rho$, we have $\varphi_A^M(I,T)\ge3\rho$ outside this event.

To bound the offline active weight, let $B_1,\dots,B_m$ be the values of~$w^*$ on the independent amplified samples entering the last averaging step. By~\cref{lem:amplification}, they are independent with $0\le B_i\le2M$ and $\E[B_i]\le1$. The offline schedule of~\cref{lem:copies}\ref{it:offline}, with weights divided by~$m$, gives $w^*(I,T)\le\bar B$ for the average $\bar B=m^{-1}\sum_iB_i$, which has expectation at most~$1$. Hoeffding's inequality~\cite{Hoeffding63} therefore implies
\begin{equation*}
  \Pr\bigl[w^*(I,T)>2\bigr]\le\Pr\bigl[\bar B-\E[\bar B]>1\bigr]\le\exp\biggl(-\frac{2m}{(2M)^2}\biggr)=\exp\bigl(-2(k+1)^2\cdot4^k\bigr)\,.
\end{equation*}
Both failure probabilities are at most $\exp(-2\cdot4^k)$. Hence every deterministic~$A$ satisfies
\begin{equation*}
  \varphi_A^M(I,T)\ge3\rho
  \qquad\text{and}\qquad
  w^*(I,T)\le2
\end{equation*}
with probability at least $1-2\exp(-2\cdot4^k)\ge1-2^{-4^k}\ge1-2^{-2^\rho}$, where the last inequality uses $k\ge\rho$.

We now flood~$\cD_N$ with the stream length $L=2H_N^2+2$ from the proof of~\cref{thm:quant-refined}, applied to the construction with~$2\rho$. Whenever $\varphi_A^M(I,T)\ge3\rho$ and $w^*(I,T)\le2$, the two bounds of~\cref{lem:flooding} give
\begin{equation*}
  F_A(\hat I)-\rho\cdot \OPT(\hat I)
  \ge3\rho L-\rho\bigl(\tfrac52L+H_N^2\bigr)
  =\rho(L/2-H_N^2)=\rho>0\,.
\end{equation*}
Thus the resulting finite distribution~$\hat\cD$ satisfies $\Pr_{\hat I\sim\hat\cD}[F_A(\hat I)\ge \rho\cdot \OPT(\hat I)]\ge1-2^{-2^\rho}$ for every deterministic~$A$. The parameter bounds follow from the proof of~\cref{thm:quant-refined}, with an absolute constant enlarged to account for the use of~$2\rho$.

Finally, average over $A\sim\cA$. Since~$\hat\cD$ has finite support, we may exchange the expectations over the algorithm and the instance, giving
\begin{equation*}
  \E_{\hat I\sim\hat\cD}
  \left[\Pr_{A\sim\cA}\bigl[F_A(\hat I)\ge \rho\cdot \OPT(\hat I)\bigr]\right]
  \ge1-2^{-2^\rho}\,.
\end{equation*}
Hence some instance in the support of~$\hat\cD$ satisfies the claim.
\end{proof}

\subsection{Nearly Known Instances}\label{sec:interarrival}

We prove~\cref{thm:interarrival}. All randomness in the distributions~$\cD_q$ of~\cref{sec:iteration} comes from the checkpoint indices~$K$ of the amplification steps, one for each heavy job. Such an index determines only how long after the heavy job the next job is released, and with a large bound~$H$ in~\cref{def:concatenation}, this time varies by a factor of at most $1+\eps$.

Let $(I,T)$ be a terminal instance whose~$n$ jobs have distinct release times $r_1<\dots<r_n$. Its \emph{interarrival times} are the~$n$ numbers $r_2-r_1,\dots,r_n-r_{n-1}$ and $T-r_n$, in this order, so the terminal time counts as the final arrival. For $\eps>0$, a terminal distribution is \emph{$\eps$-predictable} if it has the following two properties.
\begin{enumerate}[(a)]
\item\label{it:pred-jobs} All terminal instances in its support have the same number~$n$ of jobs, with distinct release times, the first of which is~$0$. For every $j\in[n]$, the $j$th released job has the same size and the same weight throughout the support.
\item\label{it:pred-gaps} The~$n$ interarrival times are mutually independent, and for every $j\in[n]$ there is a number $g_j>0$ such that the $j$th interarrival time lies in $[g_j,(1+\eps)g_j]$.
\end{enumerate}

\begin{lemma}\label{lem:predictable}
Let $\eps>0$, and let~$\cD$ be an $\eps$-predictable terminal distribution with a bound~$H$, which is used to form $\Avg_m(\cD)$ and $\Amp_M(\cD)$.
\begin{enumerate}[(i)]
\item\label{it:pred-avg} For every positive integer~$m$, the distribution $\Avg_m(\cD)$ is $\eps$-predictable.
\item\label{it:pred-amp} If $H\ge1/(2\eps)$, then $\Amp_M(\cD)$ is $\eps$-predictable for every integer $M\ge1$.
\end{enumerate}
\end{lemma}

\begin{proof}
Let~$\bar I$ be the concatenation of~$m$ copies of~$\cD$, formed from independent samples $(I_1,T_1),\dots,(I_m,T_m)$, and let~$\tau$ be its end. We first show that the distribution of $(\bar I,\tau)$ is $\eps$-predictable.

Copy~$i$ is obtained from~$I_i$ by mapping each release time~$r$ to $u_{i-1}+s_ir$ and multiplying each size by~$s_i$, where the factor~$s_i$ does not depend on the samples. Since the first job of~$I_i$ is released at time~$0$, the first job of copy~$i$ is released at~$u_{i-1}$, and all jobs of copy~$i$ are released during $[u_{i-1},u_i)$ by~\cref{lem:copies}\ref{it:geometry}. Hence the release times of~$\bar I$ are distinct, the first of them is $u_0=0$, and the jobs of~$\bar I$ in release order are the jobs of copies $1,\dots,m$ in release order, with sizes and weights that are the same for all samples. This is~\ref{it:pred-jobs}.

For~\ref{it:pred-gaps}, note that the last interarrival time of $(I_i,T_i)$, multiplied by~$s_i$, is the time from the last release of copy~$i$ to~$u_i$, and that~$u_i$ is the first release time of copy~$i+1$ if $i<m$ and equals~$\tau$ if $i=m$. The interarrival times of $(\bar I,\tau)$ are therefore those of $(I_1,T_1)$ multiplied by~$s_1$, followed by those of $(I_2,T_2)$ multiplied by~$s_2$, and so on. They are mutually independent because the samples are independent, and an interarrival time of $(I_i,T_i)$ in $[g,(1+\eps)g]$ becomes one in $[s_ig,(1+\eps)s_ig]$.

Dividing all weights by~$m$ preserves both properties, which proves~\ref{it:pred-avg}. For~\ref{it:pred-amp}, let $m=k=2M+1$. Amplification adds to~$\bar I$ the job~$J$ with release time~$\tau$, size~$V$, and weight~$2M$, divides all weights by~$\beta$, and replaces the terminal time~$\tau$ by~$h_K$. All jobs of~$\bar I$ are released before~$\tau$, so~$J$ is released last, and its size and weight are the same for all samples. The interarrival times of $(\bar I,\tau)$ remain interarrival times, the last of them now ending at the release of~$J$, and there is one new interarrival time $h_K-\tau=V+S_K$. It is independent of the others because~$K$ is independent of the copies. By~\cref{def:concatenation}, we have $V=S_0=S_1+Hs_1=(2H+1)S_1+2H$, and therefore $S_1\le V/(2H)\le\eps V$. Since $0=S_k\le S_K\le S_1$, the new interarrival time lies in $[V,(1+\eps)V]$.
\end{proof}

\begin{proof}[Proof of \cref{thm:interarrival}]
Fix $\eps>0$ and $\rho\ge1$. As in the proof of~\cref{thm:probability}, use the construction of~\cref{sec:iteration} with~$2\rho$ in place of~$\rho$ and the bounds from the proof of~\cref{thm:quant-refined}, except that now $H_0=\max\{3,\lceil1/(2\eps)\rceil\}$. That proof uses only that~$H_0\ge3$ is an integer and a bound for~$\cD_0$, so $H_q=H_0^{(6km)^q}$ is a bound for~$\cD_q$, and the amplification of round~$q$ uses the bound $H_{q-1}\ge1/(2\eps)$.

The single-job distribution~$\cD_0$ is $\eps$-predictable, since its job is released at time~$0$ and its only interarrival time is~$1$. By~\cref{lem:predictable} and induction on~$q$, every~$\cD_q$ is $\eps$-predictable. Flooding~$\cD_N$ with the stream length $L=2H_N^2+2$ releases the first stream job at the terminal time~$T$, so the last interarrival time of $(I,T)$ becomes the time before the first stream job, and the further stream jobs follow at the fixed spacing $1/(2M)$. All stream jobs have the same size and weight. The resulting random instance~$\hat I$ therefore has the form required in~\cref{thm:interarrival}. Its sizes and weights in release order are fixed, its first job is released at time~$0$, and its interarrival times are independent, each with finitely many values in an interval $[g,(1+\eps)g]$ with $g>0$.

With this choice of~$H_0$, the proof of~\cref{thm:quant-refined} at target~$2\rho$ gives $\E[F_A(\hat I)]\ge2\rho\cdot\E[\OPT(\hat I)]$ for every deterministic online algorithm~$A$. The proof of~\cref{thm:probability} at target~$\rho$ uses the same distribution and gives $\Pr[F_A(\hat I)\ge\rho\cdot \OPT(\hat I)]\ge1-2^{-2^\rho}$. The parameter bounds change only through~$H_N$, for which now $\log\log H_N=N\log(6km)+\log\log H_0$. Hence they hold with a constant~$C$ that depends only on~$\eps$, and~$C$ can be chosen independently of~$\eps$ whenever $\eps\ge2^{-2^{2^\rho}}$.
\end{proof}

\section*{AI Declaration}

A first version of a lower bound construction was found by GPT~6 Astra. The author then refined and simplified the proof and added the extensions in several rounds with the help of Claude Opus~5.5 and Claude Fable~5.2, which were also used to revise the text and create the figures. The author has checked all statements and proofs and takes full responsibility for the content of this paper.

\begin{refcontext}[sorting=nyt]
\printbibliography
\end{refcontext}

\end{document}